# Interface and film engineering for improved spin-charge interconversion in sputtered GeTe nanodevices

S. Teresi[1], T. Frottier[1], M. Culot[1], W. Savero-Torres[1,3], P. Sgarro[1], M. Wissmann[1], A. Kandazoglou[1], F. Osana[1], J. Lagrave[2], D. Térébénec[2], J. Paterson[2], A. Oysel Mestre[2], O. J. Mouawad[2], E. Chevilley[2], N. Bernier[2], P. Noé[2], M. Bibes[3,4], F. Hippert[2], J.-P. Attané[1,3], L. Vila[1,3] and M. Cosset-Chéneau[1]

[1]*Univ Grenoble Alpes, CEA, CNRS, Spintec, 17 Avenue des Martyrs, 38054 Grenoble Cedex, France*
[2]*Univ Grenoble Alpes, CEA, CNRS, Leti, 17 Avenue des Martyrs, 38054 Grenoble Cedex, France*
[3]*Nellow, 17 Avenue des Martyrs, 38054 Grenoble Cedex, France*
[4]*Laboratoire Albert Fert, CNRS, Thales, U. Paris-Saclay, 1 Avenue Augustin Fresnel, 91767 Palaiseau Cedex*

Keywords: ferroelectric Rashba semiconductors, GeTe, spin-charge interconversion, spin-orbit logic, nanodevices, magnetron sputtering

# ABSTRACT

Spin-logic devices promise non-volatile logic-in-memory computing with ultra-low power consumption, offering a route to circumvent the von Neumann bottleneck. Magnetoelectric and ferroelectric spin-orbit (MESO/FESO) devices, however, require spin-charge interconversion materials that combine a large output signal with properties that survive nanoscale integration. Tellurides are particularly fragile in this respect, as processing readily introduces defects that degrade their spin-charge interconversion properties. Here we demonstrate room-temperature spin-charge interconversion in patterned FESO nanoscale devices based on sputtered GeTe, with feature sizes down to 100 nm. By introducing a Ti interfacial layer between the CoFe electrode and GeTe, and by using GeTe films of improved crystalline quality, we increase the interconversion output signal from 3 mΩ to 24 mΩ, exceeding values reported for platinum at comparable device size. These results establish large-scale-deposited GeTe as a viable material platform for spin-logic devices and identify interface quality and film crystallinity as key levers for optimizing their performance.

## Introduction

Since the discovery of giant magnetoresistance [1,2], tunnel magnetoresistance [3], and their application in spin-transfer-torque magnetic random access memory (STT-MRAM) [4,5], the magnetic tunnel junction (MTJ) has been the leading spintronic technology. In 2011, spin-orbit torque (SOT) switching was demonstrated [6,7], enabling magnetization reversal via the spin current generated by passing a charge current in an adjacent spin-Hall material and opening the way to SOT-MRAM [6,7]. This triggered an intensive search for materials producing high spin-charge interconversion efficiency, as required for these applications [8-14].

Interest in spin-charge interconversion was renewed in 2019, when Intel proposed a spintronic device for in-memory and spin-logic computing: the magnetoelectric spin-orbit (MESO) device [15], with the potential to overcome the von Neumann bottleneck [16]. In the MESO device, information is stored in the polarization state of a magnetoelectric material. The reading unit consists of a ferromagnet coupled to a spin-Hall or Rashba material, while the writing unit is a magnetoelectric material (ME) capable of switching the ferromagnet's magnetization without an external field. The minimum voltage required for ME switching is on the order of 100 mV [17-21], so research has focused on materials with larger spin-Hall or Rashba-Edelstein effects, since a correspondingly large interconversion signal is needed to generate this voltage and allow MESO devices to be cascaded [8-14]. A first demonstration of controlling spin-charge interconversion via a magnetoelectric component was recently reported [22].

To alleviate MESO's need for a multiferroic material to switch magnetization, an alternative was proposed the following year: the ferroelectric spin-orbit (FESO) device [23], in which a ferromagnetic electrode injects a spin current into a system displaying spin-orbit coupling tunable by ferroelectricity. Here, the sign of the spin-charge interconversion is controlled directly by the ferroelectric polarization direction, removing the need for magnetization switching altogether. FESO thus merges the reading and writing functions into a single material combining spin-orbit coupling (SOC) and ferroelectricity and is expected to serve the same spin-logic and in-memory computing roles as MESO.

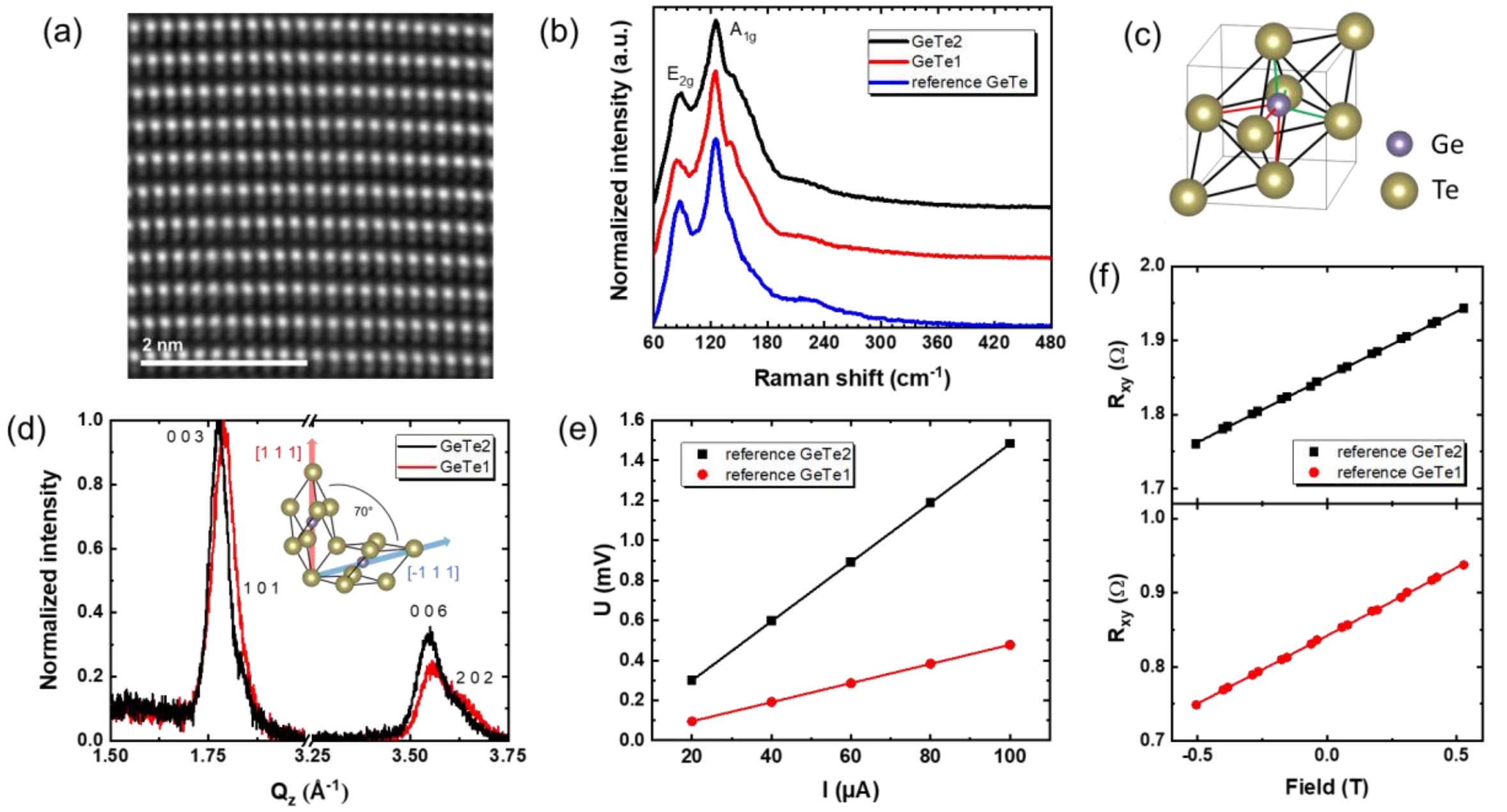

Figure 1. (a) High-angle annular dark-field scanning transmission electron microscopy (HAADF-STEM) image of a reference GeTe thin film grown on $SiO_2$, viewed along the ⟨2 1 0⟩ zone axis, revealing unequally spaced Ge and Te planes parallel to the film surface. The image was denoised by selectively masking noise-related frequency components in Fourier space, followed by inverse Fourier transformation. (b) Raman spectra of the reference film shown in (a) and the GeTe1 and GeTe2 films, showing the characteristic E2g and A1g modes of rhombohedral GeTe. (c) Schematic of the rhombohedral α-GeTe unit cell represented within a pseudo-cubic unit cell elongated along its body diagonal. Te atoms (yellow) occupy the corners and face-centered positions of the pseudo-cubic cell, while the Ge atom (purple) is displaced from the centrosymmetric position along the [111] direction. This off-centering distorts the ideal octahedral Ge–Te coordination, resulting in three short (green) and three long (red) Ge–Te bonds, and gives rise to the spontaneous polarization associated with the ferroelectric distortion. (d) X-ray diffraction (XRD) patterns acquired on GeTe2 and GeTe1 films deposited using two different deposition conditions and showing the change in crystalline quality between both films (for details see text). (e) Voltage-current characteristics of the 100 nm thick GeTe films for GeTe1 and GeTe2. (f) Ordinary Hall effect measurements for the same reference films.

| Sample | Resistivity [Ω cm] | Carrier density [$cm^{-3}$] | Carrier mobility [$cm^2\ V^{-1}\ s^{-1}$] |
|---|---|---|---|
| GeTe1 | $2.39 \pm 0.05 \times 10^{-4}$ | $3.52 \pm 0.09 \times 10^{20}$ | $74.25 \pm 0.42$ |
| GeTe2 | $7.40 \pm 0.42 \times 10^{-4}$ | $3.80 \pm 0.10 \times 10^{20}$ | $22.24 \pm 1.87$ |

Table 1. Transport properties calculated from measurements shown in Figure 1e and 1f for 100 nm-thick GeTe1 and GeTe2 reference samples.

In addition, using ferroelectric polarization rather than magnetization as a state variable should enable higher operation frequency and strongly limit cross-talk in dense arrays. As for MESO, cascading FESO devices in a logic circuit requires the SOC component to generate a transverse output voltage of about 100 mV. Identifying ferroelectric systems with large spin-charge interconversion is therefore a central challenge for FESO.

GeTe is a well-established ferroelectric Rashba semiconductor in which spin-charge conversion can be controlled by ferroelectricity at room temperature [24,25]. It can be grown with strong preferential texture on a large wafer scale using an $Sb_2Te_3$ seed layer, and is widely employed in GeTe/$Sb_2Te_3$ superlattices for phase-change memory devices [26-29] (Figure 1). In this work, we focus on fabricating spin-charge interconversion nanodevices based on GeTe grown by sputtering on an industrial deposition platform.

We demonstrate room-temperature interconversion in nanopatterned FESO devices made of GeTe film grown by sputtering onto a TiN layer, and show that the quality of the interface between the ferromagnetic top electrode and GeTe, together with the crystalline quality of the GeTe film, strongly affects the interconversion signal.

We first present the structural and electrical properties of the sputtered GeTe thin film, then describe the fabrication process used to integrate GeTe into the FESO device. We next show via HAADF-STEM how nanofabrication choices can be tuned to control the structural quality of the GeTe layer integrated in a device. Finally, we show how the spin-charge interconversion signal depends on GeTe quality and interface properties.

## Material growth and characterization

GeTe samples with thickness ranging between 5 and 20 nm were deposited by magnetron sputtering in an industrial cluster tool at 300°C on a 10 nm thick polycrystalline TiN layer deposited on 200 mm Si(100) wafer capped with a 500 nm-thick thermal silicon oxide as a bottom insulating layer. Following GeTe deposition, a 5 nm thick $Sb_2Te_3$ capping layer was deposited in situ on top of GeTe films to prevent any possible surface oxidation before device fabrication.

Two sets of GeTe growth conditions were investigated, yielding films with different compositions and properties, hereafter denoted GeTe1 and GeTe2. The different growth conditions (to be detailed in a future publication), lead to different crystalline quality. Raman spectra from both GeTe samples (Figure 1b) exhibit modes characteristic of the polar rhombohedral α-GeTe phase, whose structure is shown in Figure 1c. These modes closely match those measured in a reference film [30]. A cross-sectional HAADF-STEM image of a reference film (Figure 1a) shows stacked Ge and Te atomic planes with relative off-centering consistent with an out-of-plane polarization (so called GeTe(111)). The main difference between GeTe1 and GeTe2 is evidenced by the out-of-plane XRD patterns shown in Figure 1d. For the GeTe2 films conditions, all diffraction peaks can be indexed as (0 0 *l*) peaks (hexagonal indexation) of the rhombohedral α-GeTe phase indicating that the c-axis of GeTe crystallites is along the normal to the substrate's plane. On the other hand, in case of GeTe1 film, additional shoulders are visible close to the position of the (003) and (006) Bragg peaks of c-axis oriented α-GeTe(111). The latter doublets can be indexed as (003)/(101) and (006)/(202) peaks. They indicate the presence of a non-negligible fraction of rotated ferroelectric domains with their axis tilted from the normal to the substrate's surface (70° to the normal). Therefore, the main difference between both deposition conditions stands in the orientation of the ferroelectric domains of α-GeTe crystallites. We will show hereafter that such structural difference can play a major role on spin-charge conversion efficiency.

To characterize the transport properties associated with the two growth conditions, Hall-effect measurements were performed in a van der Pauw geometry on 100 nm-thick GeTe1 and GeTe2 reference films deposited on 500 nm-thick $SiO_2$ film on Si(100) substrates (Figure 1f). The results are summarized in Table 1. The two films exhibit similar carrier concentrations of approximately 3.5-3.8 × $10^{20}$ $cm^{-3}$, whereas their carrier mobilities differ by more than a factor of three. Accordingly, their resistivities differ substantially, with values of (2.39 ± 0.05) × $10^{-4}$ and (7.40 ± 0.42) × $10^{-4}$ Ω.cm for GeTe1 and GeTe2, respectively. These values are consistent with those previously reported for epitaxially grown GeTe films [30]. The temperature dependence of the resistivity further supports metallic-like transport in these GeTe films (see Supplementary material S1).

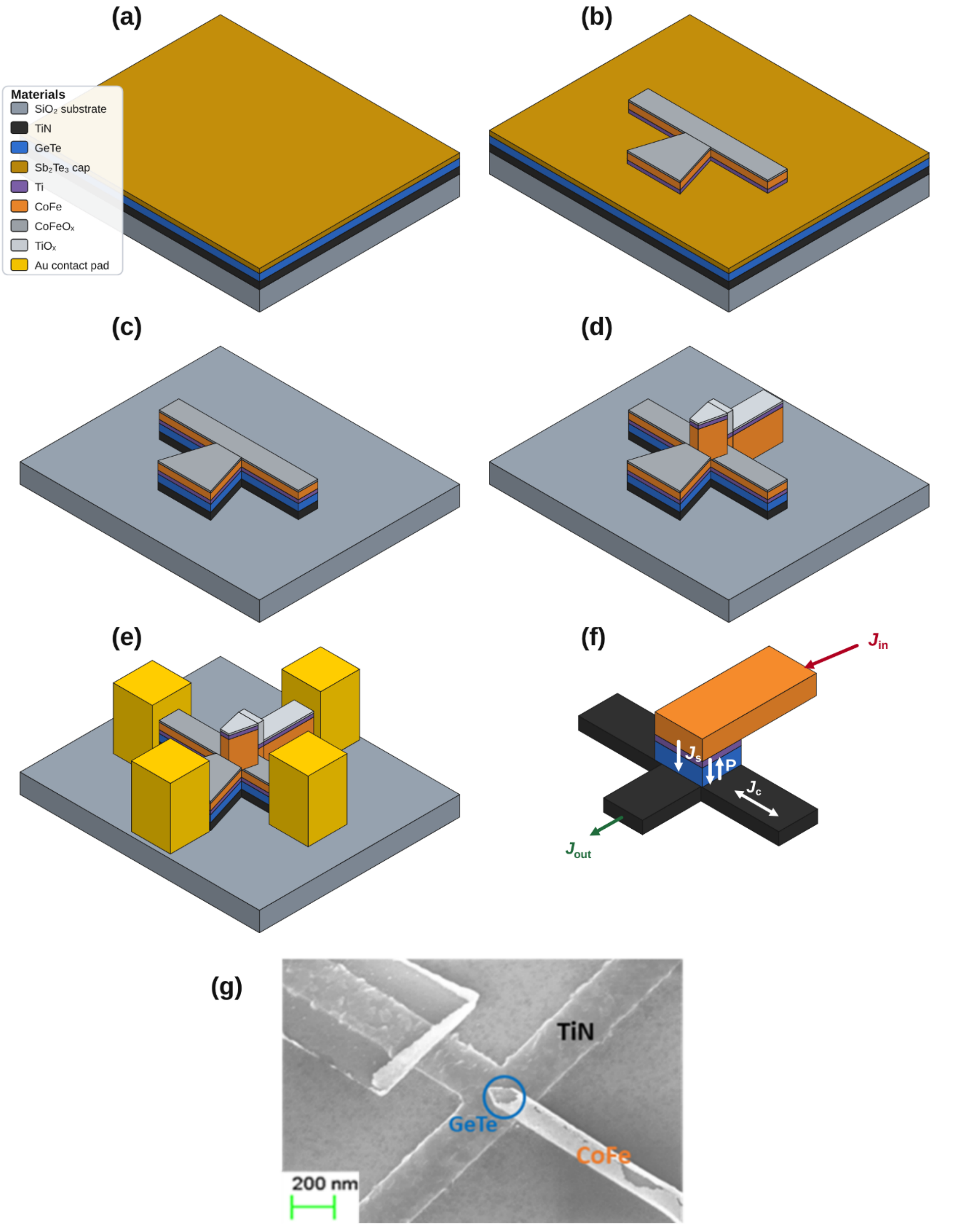

Figure 2: Nanofabrication process flow. (a) Starting stack, consisting of an $Sb_2Te_3$-capped GeTe layer grown on TiN. (b) T-shaped hard mask deposited after the first e-beam lithography step. (c) T-shaped device on the $SiO_2$ substrate obtained after etching. (d) Device after the second e-beam lithography step and deposition of the thick ferromagnetic electrode. (e) Final optical lithography step defining the gold contact pads. (f) Schematic of the completed device in the spin-to-charge configuration. (g) Scanning electron microscopy (SEM) image of a completed 250 nm wide device.

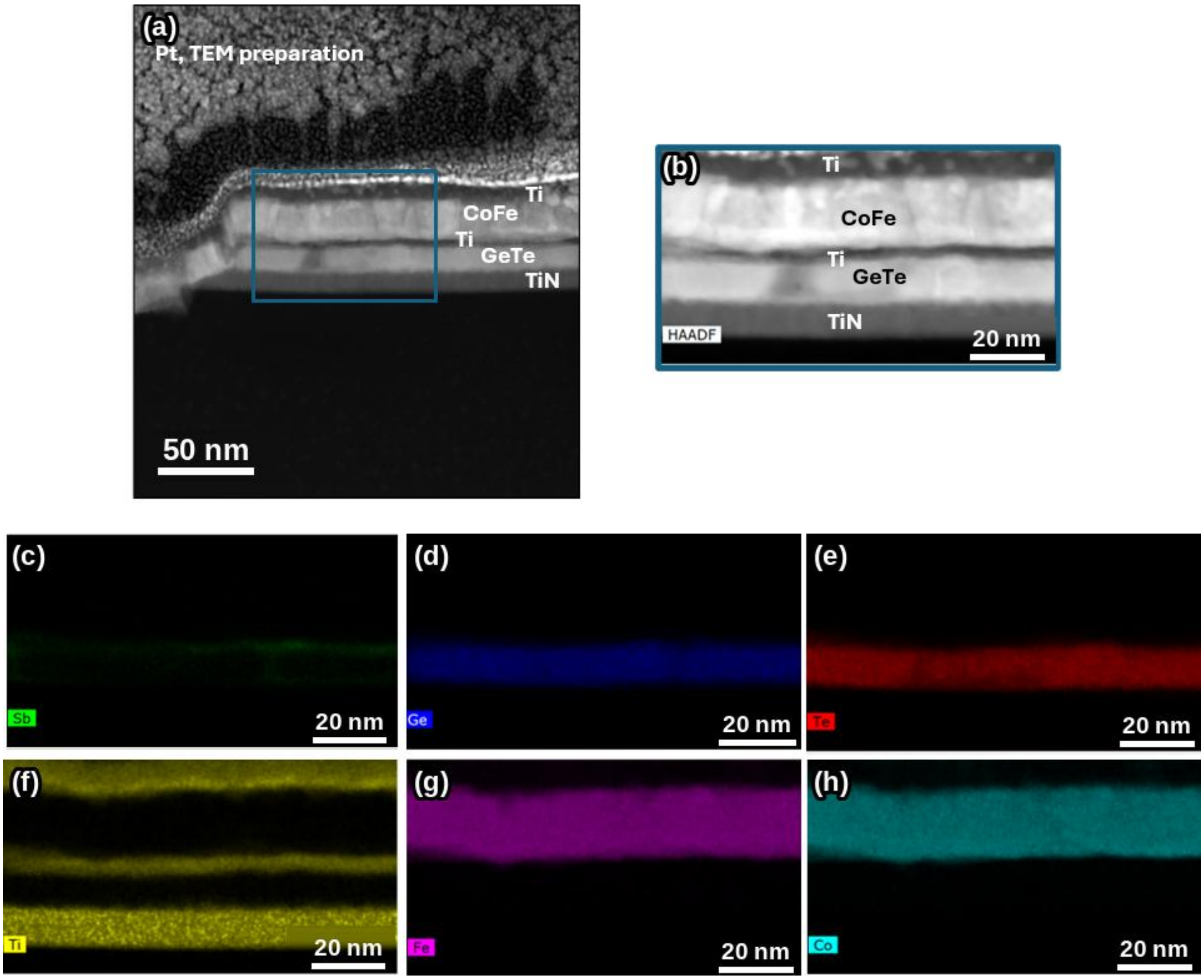

Figure 3: Structural and chemical analysis of a Ti-interlayer device (GeTe1 grown on TiN), measured on a 250 nm device. (a) STEM cross-section of the whole stacking, with the layer sequence indicated; the Pt layer is deposited for transmission electron microscopy (TEM) lamella preparation and is not part of the device. (b) HAADF-STEM image at higher magnification of the region highlighted in (a). (c, d, e) STEM energy-dispersive X-ray spectroscopy (EDX) maps of Sb, Ge and Te. Ge and Te together delimit the GeTe interconversion layer, while the weak residual Sb signal at its upper surface corresponds to unetched remnants of the $Sb_2Te_3$ capping layer. (f) EDX map of Ti, resolving the three distinct Ti-bearing layers of the stack: the TiN bottom electrode, the Ti interlayer beneath the ferromagnet, and the Ti capping. (g, h) EDX maps of Fe and Co, delimiting the CoFe ferromagnetic electrode. Neither element is resolved within the GeTe, confirming that the Ti interlayer suppresses intermixing between the ferromagnet and the interconversion layer. The complete elemental set, including O, N and Si, is given in supplementary material S3.

## Spin orbit read out device nanofabrication

The nanofabrication process used for the GeTe-based FESO devices is summarized in Figure 2. The starting stack is TiN(10 nm)/GeTe(10 nm)/$Sb_2Te_3$(5 nm) deposited on a 500 nm thick $SiO_2$ layer on a Si substrate (Figure 2(a)). The TiN layer serves as electrical contact to the active part of the device, that is the FM-covered GeTe area, and can additionally act as a back-electrode for switching the ferroelectric polarization of the GeTe layer. The $Sb_2Te_3$ capping layer protects GeTe from oxidation on exposure to atmosphere and is removed during the processing.

A T-shaped structure was first patterned on the blanket stack by electron beam lithography (EBL) (Figure 2(b)). The $Sb_2Te_3$ capping layer was then removed by soft $Ar^+$ sputtering, and a Ti (or $TiO_x$) layer (1-2 nm) was deposited by electron beam evaporation, followed by a CoFe (5 nm) layer serving both as a hard mask and as a seed for the ferromagnetic (FM) spin injection layer deposited subsequently.

The uncovered TiN/GeTe/$Sb_2Te_3$ was subsequently removed by reactive ion etching (RIE) using an $SF_6/CH_2F_2$ gas mixture (Figure 2(c)), leaving only the patterned T-shape. In a second EBL step, a single line was patterned to partially overlap the center of the T-shape. The FM electrode (20 nm) was then deposited, after removing the oxidized surface of the CoFe seed layer by ion beam etching (Figure 2(d)). A final ion beam etching leaving TiN only as the branches of the T-shape, leaving GeTe only underneath the FM electrode. Macroscopic electrical contacts were finally defined by optical lithography followed by Ti/Au evaporation (Figure 2(e)).

The resulting device, used for all the interconversion measurements reported below, is shown schematically in Figure 2(f); an SEM image of a completed 250 nm device is shown in Figure 2(g).

## STEM analysis

To assess the interfacial quality and, more generally, the validity of the nanofabrication process, we acquired cross-sectional STEM images of our devices. In contrast with our previous work on $Sb_2Te_3$ [31], the CoFe electrode thickness was chosen such that it does not generate sufficient strain to provoke significant degradation of the structural quality of GeTe film. This is confirmed by the absence of significant cracking in our STEM images (Figure 3(b) and Figure S3). Reducing the CoFe thickness also improved the device yield, from 5% to 15%.

The images show good interface quality, with no detectable intermixing between layers (Figure 3). The GeTe thickness is slightly larger than 10 nm, indicating that although nearly all the $Sb_2Te_3$ capping layer is removed during the fabrication process, some residue may remain. The Ti layer is continuous across the interface, confirming its role as a diffusion barrier between the atomic species of the spin-charge interconversion material from those of the FM: it acts both as an intermixing barrier and as a layer favoring spin injection [32].

## Electrical measurements

All spin-charge interconversion measurements were performed at room temperature. A bias current is applied across the CoFe/Ti/GeTe interface, injecting a spin current into GeTe with a spin polarization set by the CoFe magnetization direction (Figure 2(f)). This spin current is converted into a transverse charge current in GeTe, which we detect as a transverse voltage, or as a transverse resistance when normalized by the bias current (Figure 4).

To read out the interconversion signal, an in-plane magnetic field is swept along the easy axis of the CoFe electrode, switching its magnetization between two saturated states. This reverses the sign of the injected spin polarization and, correspondingly, the sign of the transverse voltage (Figure 4(a)-(c)). Table 2 summarizes the results obtained for the different stacks shown in Figure 4. We define the interconversion output signal as $2\Delta R_{xy}$, the difference in transverse resistance measured at positive versus negative field [33]. Local configurations of this kind are known to be sensitive to spurious Hall contributions [34-36]. In particular, the planar Hall effect arising from a misalignment between the magnetization and the current direction in the FM electrode [35], and the anomalous Hall effect generated in the FM [37] can contribute to the transverse signal. The finite baseline visible in Figure 4 originates from the misalignment of the CoFe electrode and cross-shaped structure fabrication [38], and does not affect the extracted output signal, since $2\Delta R_{xy}$ is defined as a difference between the two saturated states and is therefore insensitive to a constant offset.

Measurements were performed using a conventional lock-in technique: an AC bias current of 50-100 µA at a frequency of 133 Hz was applied with a current generator, and the resulting transverse voltage was recorded with a lock-in amplifier.

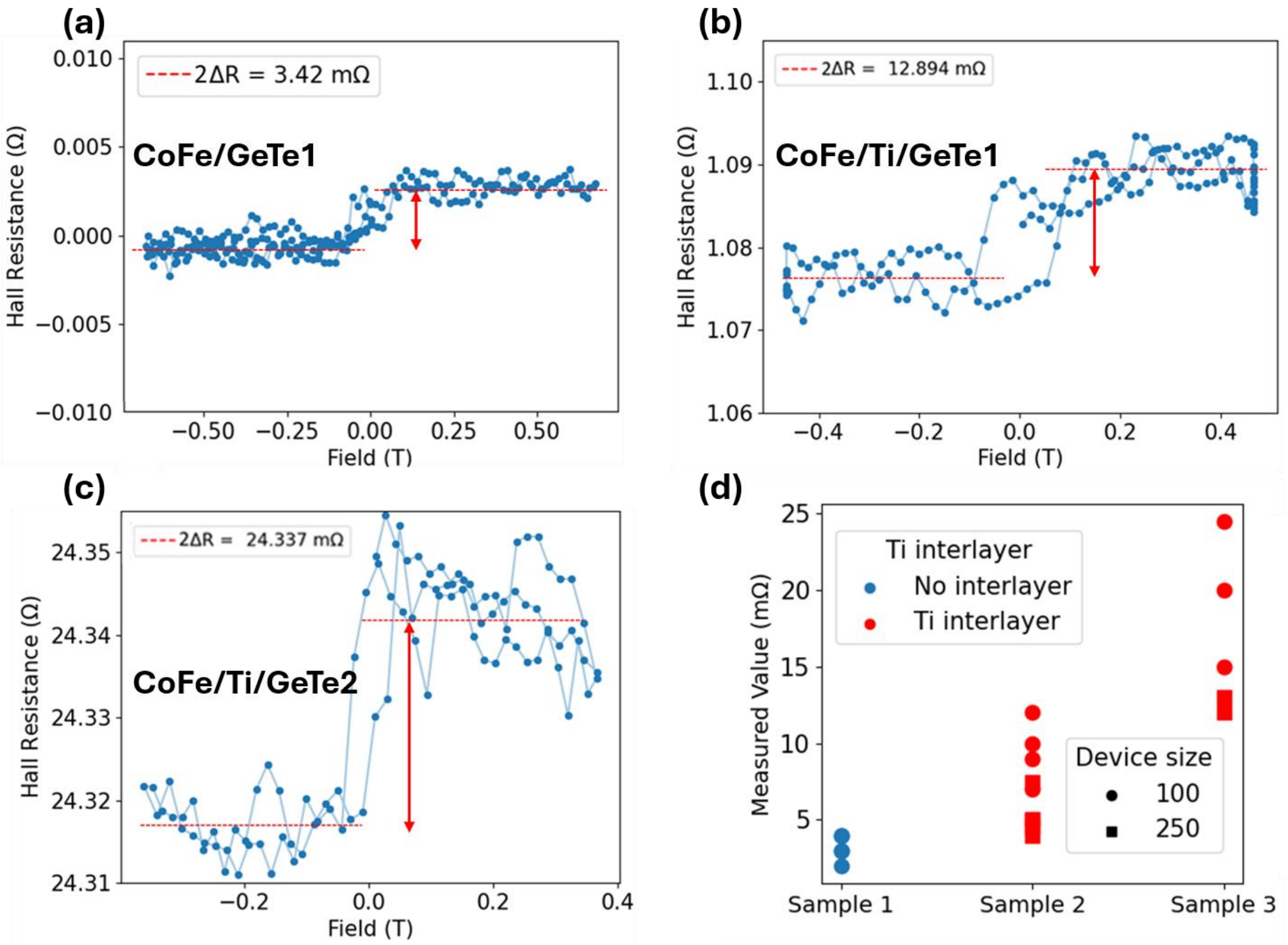

Figure 4: Spin-to-charge conversion measured at room temperature. The x axis is the applied in plane magnetic field which switches the magnetization of the ferromagnet; the y axis is the transverse direction along which the resistance is measured in the Hall configuration. (a) Sample 1, GeTe/CoFe, without interfacial layer. (b) Sample 2, GeTe1/Ti/CoFe. (c) Sample 3, GeTe2/Ti/CoFe. The baseline resistance offset in (a)-(c) originates from a slight misalignment of the ferromagnetic electrode with respect to the channel and does not affect the extracted signal (see main text). (d) Output signal $2\Delta R_{xy}$ reported for the three sample conditions, at feature sizes of 100 nm and 250 nm.

| Sample | Figure 4 panel | Stack | $2\Delta R_{xy}$ at 100 nm [mΩ] |
|---|---|---|---|
| 1 | (a) | CoFe(20)/**GeTe1(10)/**TiN(10) | 3.2 ± 0.8 (N = 5) |
| 2 | (b) | CoFe(20)/**Ti(2)/GeTe1(10)/**TiN(10) | 9.8 ± 2.5 (N = 4) |
| 3 | (c) | CoFe(20)/**Ti(2)/GeTe2(10)/**TiN(10) | 19.8 ± 4.8 (N = 3) |

Table 2. Stack composition of the three device conditions and the corresponding interconversion output signal. Layers are listed from top to bottom and thicknesses are given in nm. Samples 1 and 2 differ only by the Ti interfacial layer, samples 2 and 3 only by the GeTe growth condition. Signals are those measured at a 100 nm feature size (Figure 4(a)-(c)). Each value is the mean of the N devices measured for that condition, and the quoted uncertainty is the standard deviation over those devices.

## Spin-charge interconversion

We fabricated three samples differing in their stack composition to disentangle the effect of the FM/GeTe interface from that of the GeTe film growth conditions on the FESO device output. Unless otherwise stated, the results below correspond to devices with a feature size of 100 nm, defined as the width of the TiN channel and of the CoFe electrode (Figure 4(a)-(c)).

Samples 1 and 2 were fabricated from a GeTe1 thin film. In sample 1, the CoFe is in direct contact with GeTe. In sample 2, a Ti interfacial layer is inserted in between CoFe and GeTe to improve the interface quality and reduce intermixing [31]. Sample 3 uses the same Ti/CoFe stack as sample 2 but grown on a GeTe2 film.

Sample 1 yields an output signal of $2\Delta R_{xy}$ = 3 mΩ (Figure 4(a)), while sample 2 gives a substantially larger signal of 13 mΩ (Figure 4(b)).
The improvement can have two origins: the Ti layer prevents CoFe from intermixing with GeTe [31], and it can additionally limit spin impedance mismatch at the interface [32]. Sample 3, processed identically to sample 2 but on GeTe2, gives an output of 24 mΩ (Figure 4(c)), nearly twice the sample 2 value. Figure 4(d) collects the output signals measured across several devices for samples 1, 2 and 3 at both 100 nm and 250 nm feature sizes. The signal increases systematically from sample 1 to sample 3, and is larger at 100 nm than at 250 nm as expected from the scaling demonstrated in Ref. [38]. Together, these results indicate a correlation between larger output signals and use of GeTe2 in the device stack. This shows that maximizing the output signal can be achieved by both scaling the devices down and jointly optimizing film growth and device design to improve spin injection. This is the first demonstration of nanofabricated devices based on sputtered GeTe, reaching spin-charge interconversion signals larger than those reported for platinum at comparable device size [38].
We also attempted to switch the ferroelectric polarization to control the sign of the spin-charge interconversion signal, as required for FESO operation and as already demonstrated in thin films [24,25]. The present device design did not allow ferroelectric switching, owing to current shunting (see supplementary material S5).

## Conclusions

We have demonstrated the nanofabrication of a FESO device based on large-scale deposited GeTe. The ability to integrate phase-change materials into nanodevices is of interest well beyond spintronics, since most of these compounds are coined as metavalent and are of high interest in thermoelectric and phase-change electronic and photonic applications [26, 28, 29, 39-42]. We have further evidenced the role of device scaling, growth conditions and spin injection barrier in optimizing the output, reporting large room-temperature interconversion signals. The origin of the interconversion, however, cannot be unambiguously assigned to either the spin Hall effect or the Rashba-Edelstein effect, since the two produce identical symmetries in our measurement geometry.
Although the signals reported here exceed those obtained with platinum at comparable device size [38], they remain well below the level required for cascading MESO and FESO devices in a logic circuit. Reaching the target output voltage of 100 mV (for MESO) corresponds to an interconversion resistance of order 100 Ω for a 1 mA bias current, several orders of magnitude above the values measured in this work. This gap is common to GeTe and to the materials with the highest spin Hall angles reported so far and closing it will require exploring new material classes [43,44].
Several routes remain to be explored to increase the interconversion signal. In-situ deposition of the interfacial Ti or $TiO_2$ layer would improve the interface quality, while a crystalline MgO barrier could optimize spin filtering and minimize spin backflow [45,46]. Recent work on orbital Hall effects in materials and at interfaces [47-49] may also contribute to improving the interconversion efficiency and could yield larger output signals when combined with the approach demonstrated here.
Finally, applying growth engineering to novel high-interconversion materials offers a path towards the output levels needed for signal cascading in MESO and FESO devices.
For FESO specifically, alternative materials combining ferroelectricity with spin-charge interconversion must be identified. GeTe alloys such as $Ge_xSn_{1-x}Te$ are a natural choice, while recent results also suggest WTe2 can be a candidate [50]. The nanofabrication process presented here is simple, reproducible and is directly transferable to these systems. Definitively demonstrating FESO operation will nonetheless require a new device design: injecting spin purely along z, in an MTJ-like geometry, would allow ferroelectric control of the spin-charge interconversion to be established at the nanodevice level.

## Acknowledgements

The authors acknowledge support from the SPEAR project (grant agreement 955671) and from the PEPR SPIN program through the ANR project Nifty (24-EXSP-0009). The authors thank the Plateforme Technologique Amont (PTA) for technical support.

## Conflict of Interest

The authors declare no conflict of interest.

# SUPPLEMENTARY INFORMATION

## S1 Temperature dependence of the sheet resistance

The sheet resistance of the full TiN/GeTe1/SbTe stack was measured in a four-probe configuration between 15 K and 290 K. It decreases monotonically on cooling, from 27.7 Ω/□ at 290 K to 24.7 Ω/□ at 15 K, confirming the metallic character of the transport, and flattens below about 50 K where the phonon contribution freezes out. Since the three layers conduct in parallel, this measurement gives the sheet resistance of the whole stack, dominated by the least resistive layer, and not the resistivity of the GeTe1 alone.

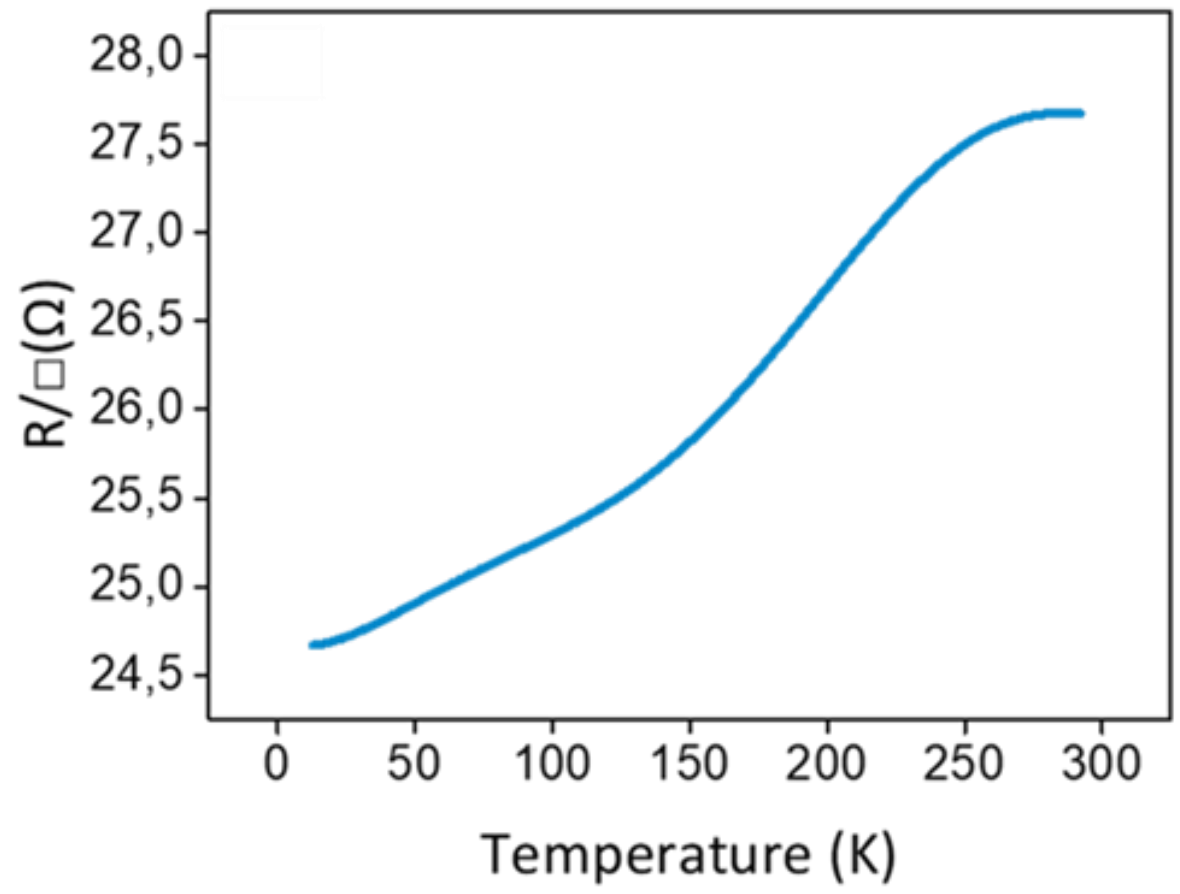

Figure S1: Sheet resistance of the full TiN/GeTe1/SbTe stack as a function of temperature, measured between 15 K and 290 K.

## S2 Direct and inverse spin-charge interconversion

The measurements shown here were performed on a 250 nm device, whose signal is smaller than that of the devices reported in the main text. The stack is TiN(10)/GeTe(10)/Ti(2)/CoFe(10)/Ti(20), with thicknesses in nm. The two configurations are related by Onsager reciprocity: exchanging the current and voltage leads reverses the sign of the signal while leaving its magnitude unchanged. This is what we measure, $2\Delta R$ = +7.6 mΩ and −8.9 mΩ, which confirms that the signal is a genuine interconversion signal.

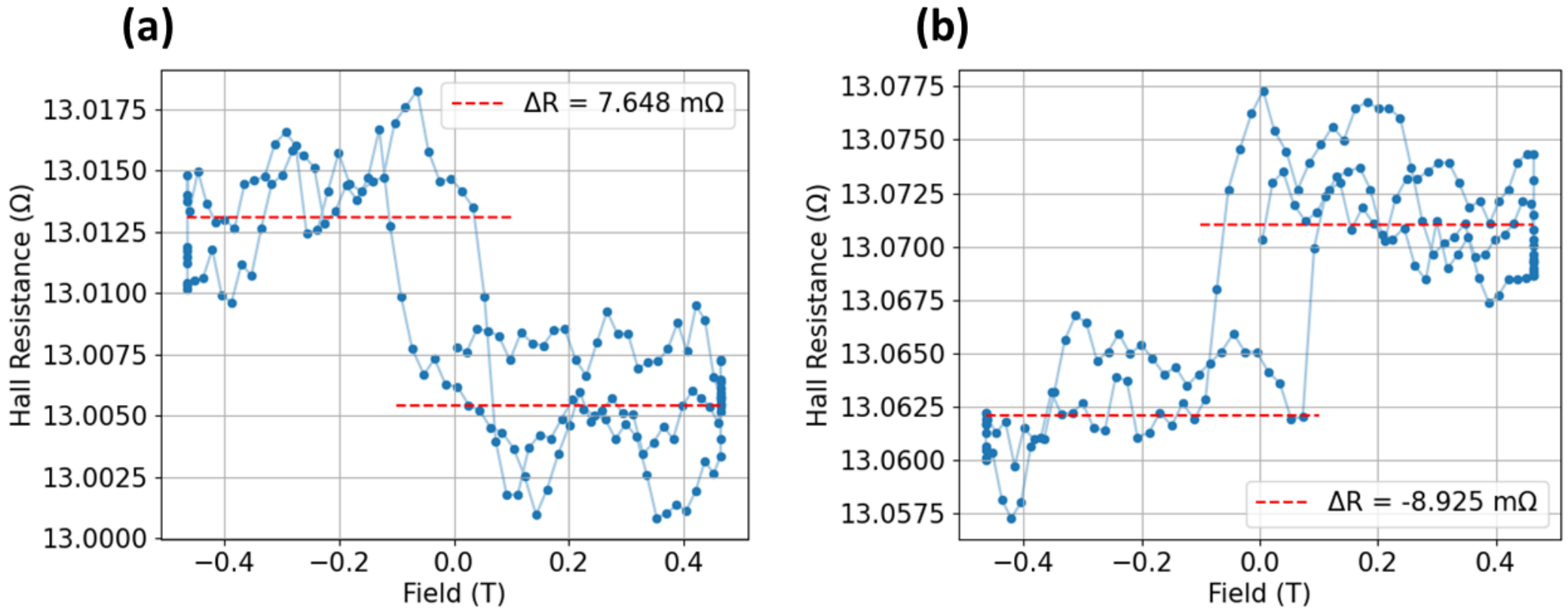

**Figure S2:** Hall resistance as a function of the in-plane magnetic field for a 250 nm device, in the direct (a) and inverse (b) configurations. The dashed lines are the average of the data above the saturation field, and their separation gives $2\Delta R$.

**S3 Detailed STEM analysis including chemical species**

This section reports the complete set of EDX-STEM maps of a 250 nm wide TiN/GeTe/Ti/CoFe device, of which Figure 3 of the main text shows a selection. Maps are given for a section of the stack far from the ferromagnetic finger (Figure S4) and for a section underneath it (Figure S5).

A weak Sb signal is detected as a thin band at the top of the GeTe, showing that the $Sb_2Te_3$ layer was not completely removed by the etching and that some unetched Sb remains at the GeTe surface, consistent with the maps of Figure 3. GeTe shows little to no Co or Fe contamination, confirming the effectiveness of Ti as a diffusion barrier at the interface. Oxygen is mainly found in the $SiO_2$ substrate and at the Ti layers; some oxygen is still visible within the GeTe, but in small quantities, possibly migrated from the Ti layer. Unlike the $Sb_2Te_3$ samples studied previously [S1], a thick CoFe layer does not degrade the GeTe crystal quality here.

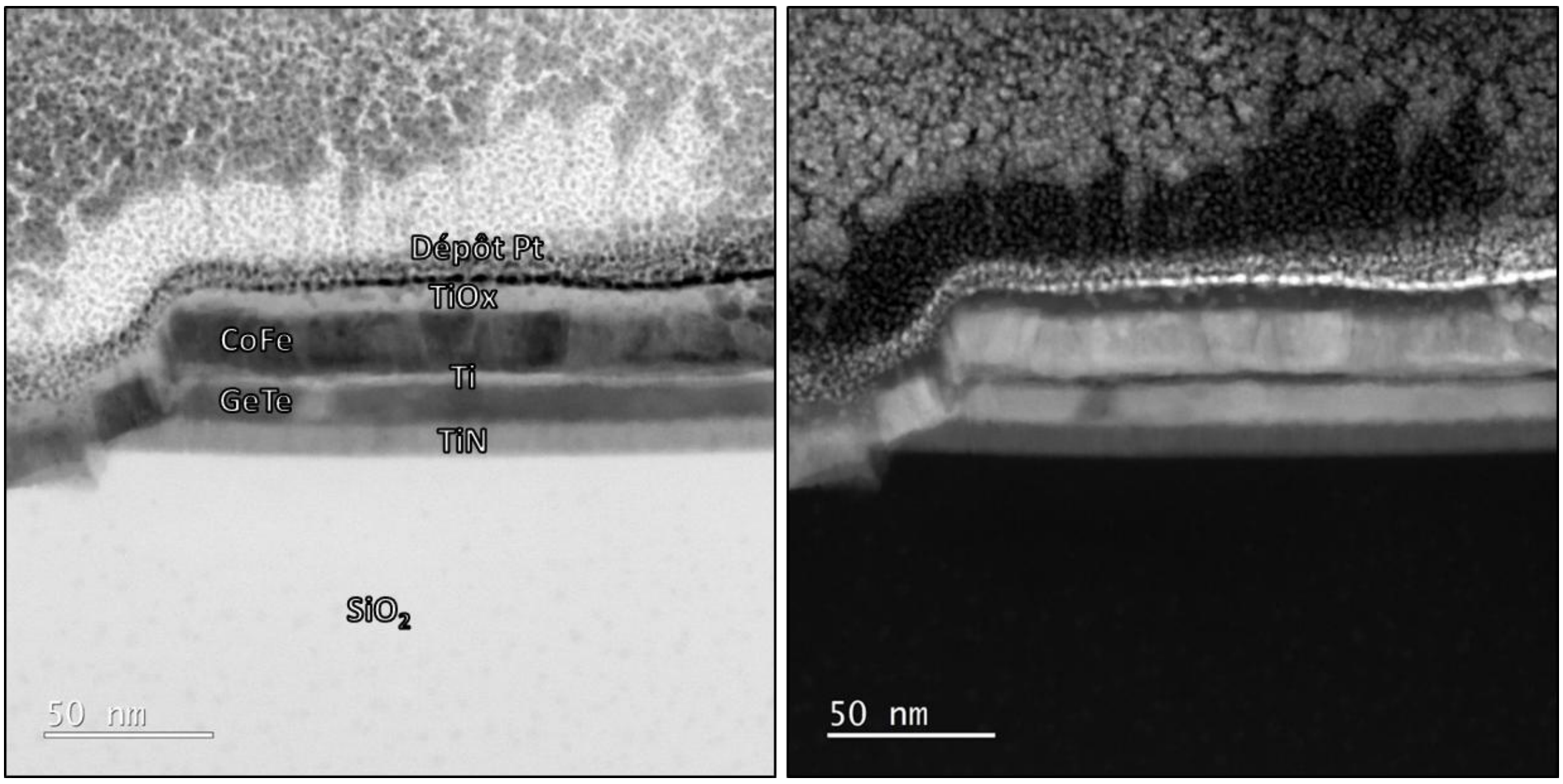

**Figure S3:** STEM cross-section of a 250 nm sample below the ferromagnet, in bright field (left) and high-angle annular dark field (right).

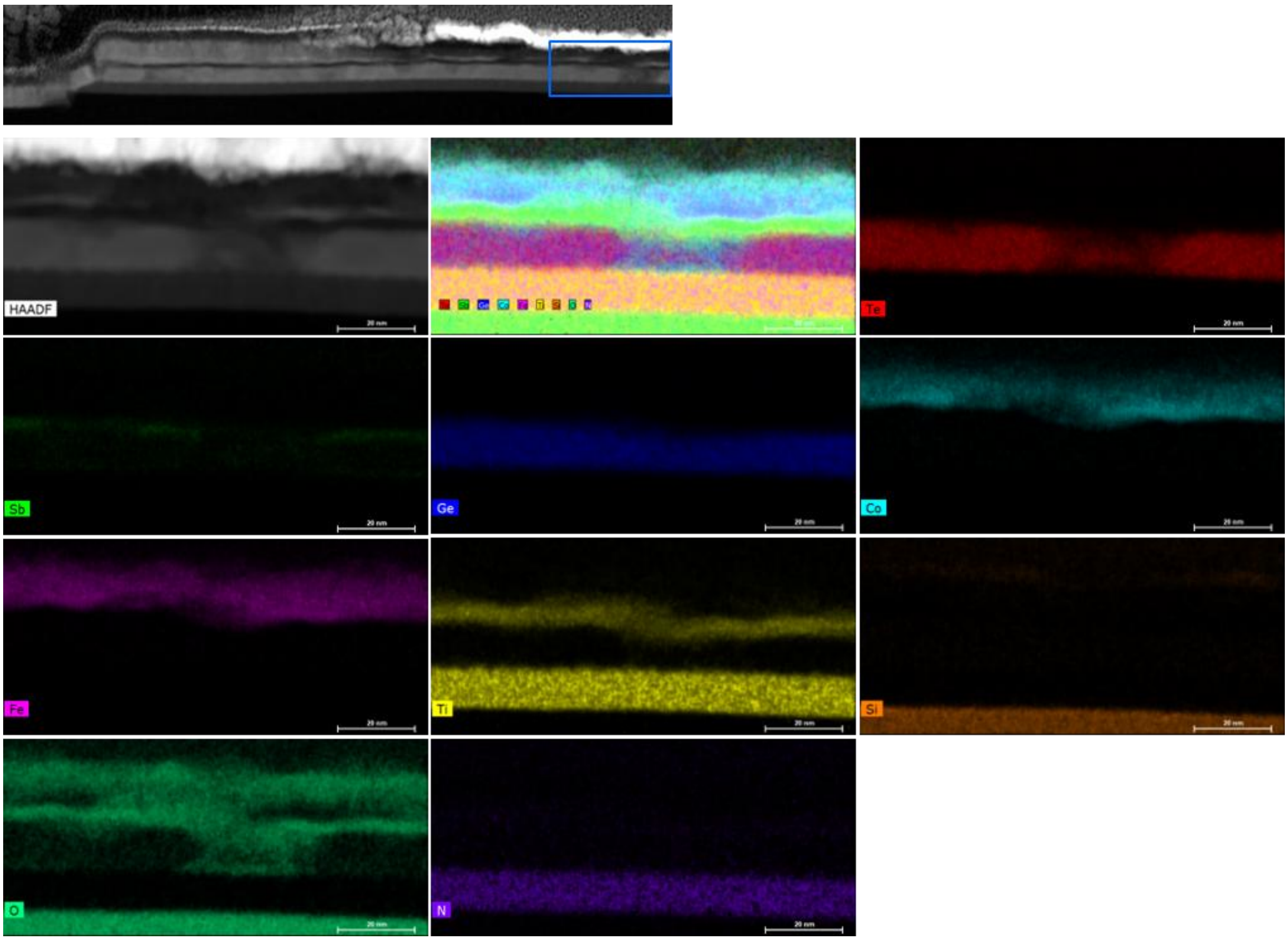

**Figure S4**: Detailed chemical composition of each layer, for a section of the device far from the spin-charge interconversion region.

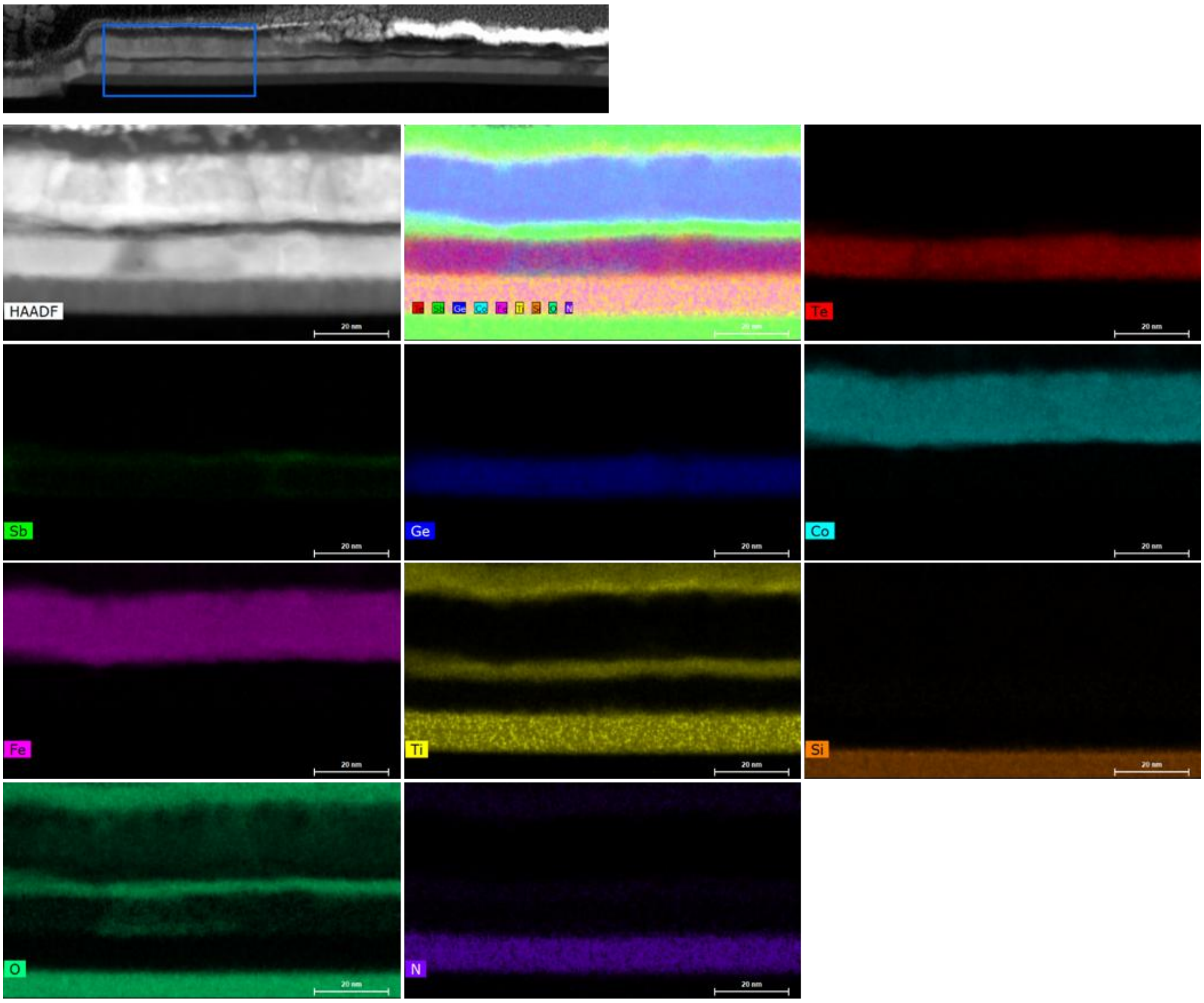

**Figure S5**: Detailed chemical composition of each layer, for a section of the device in the spin-charge interconversion region.

**S4 FEM simulations of the AHE and the spin Hall angle in GeTe**

Finite element modeling was used to estimate the anomalous Hall effect (AHE) contribution in our devices and how it scales with device size. The simulations show that the AHE accounts for only a small part of the signal measured in the main text. A CoFe thickness dependence was also computed, and identifies CoFe/GeTe combinations for which the AHE contribution is minimal.

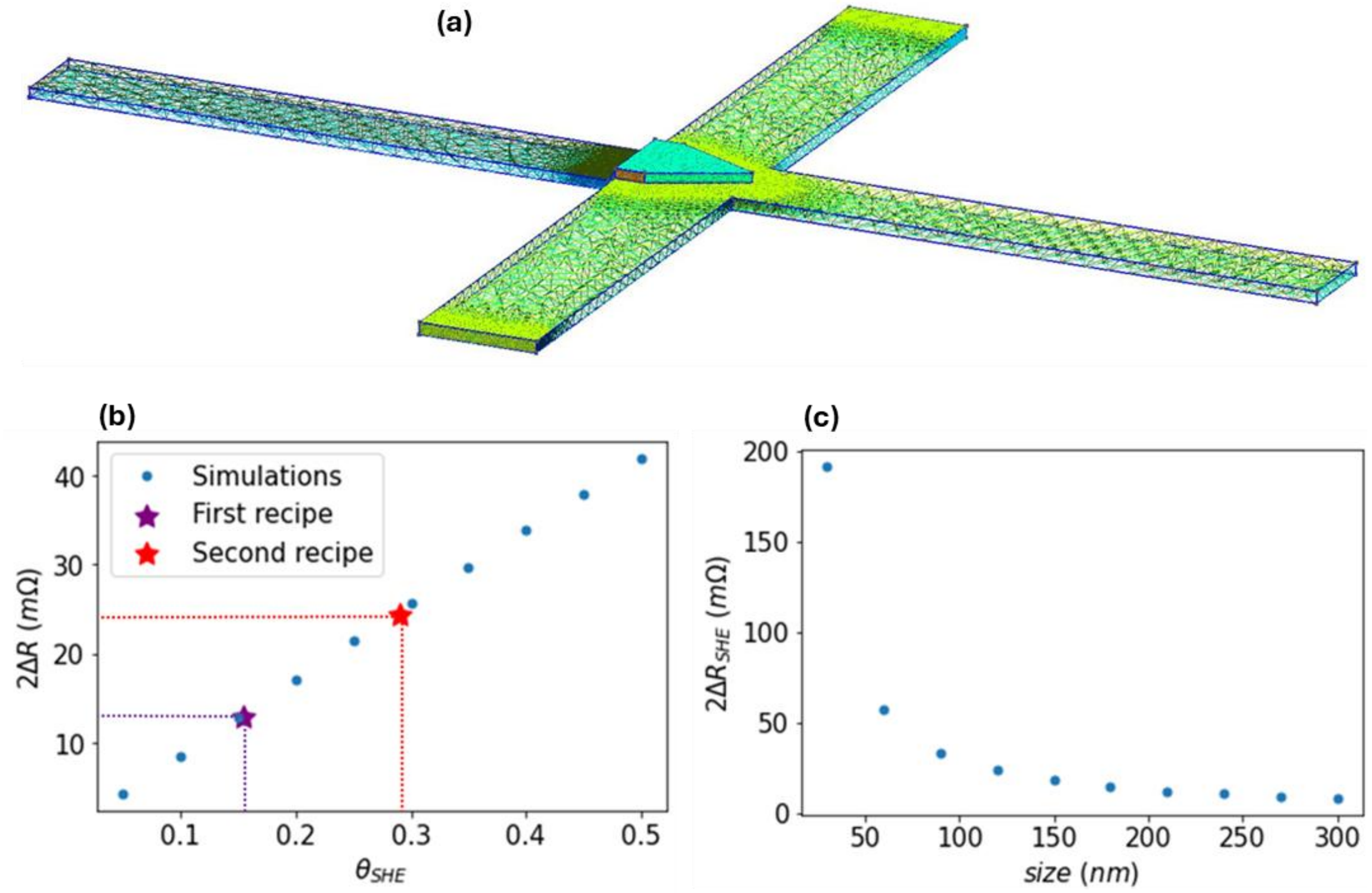

**Figure S6:** (a) Finite element modeling of a GeTe nanodevice. (b) Spin Hall angle extracted from the experimental data of the samples in Figure 4(b) and (c). (c) Estimated geometrical dependence of the output signal as a function of the device size.

The simulated device reproduces the geometry of our nanofabricated one. Assuming a spin diffusion length $\lambda$ = 9 nm taken from previous reports [S2], we extract spin Hall angles of 15% and 30% for the GeTe1 and GeTe2 nanodevices respectively. These values show that alternatives to classical heavy metals exist, and that materials grown on an industrial scale with high crystalline quality can play a relevant role in the generation of large spin-orbit torques. The simulated dependence of the output signal on device size further shows how important scaling is for increasing the output.

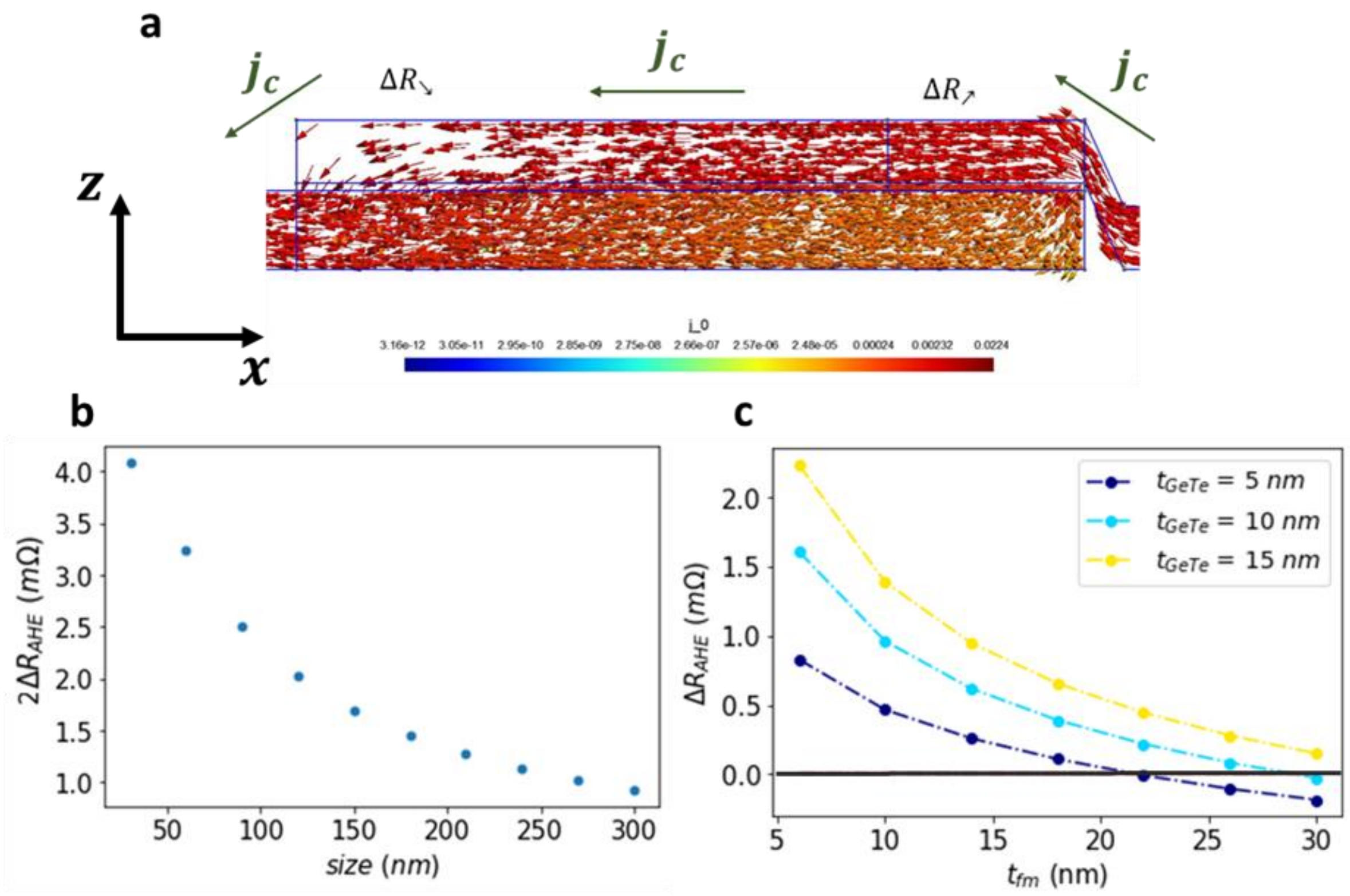

**Figure S7:** Simulation of the anomalous Hall effect (AHE) contribution to the measured signal. (a) Side view (x–z) of the interconversion region, with the ferromagnet (FM) on top injecting spin carriers into the spin-orbit coupling (SOC) material underneath. The color scale gives the local current density and the green arrows the injected charge current $j_c$. At the two ends of the FM the current turns vertically, and these two vertical paths each produce an AHE contribution, $\Delta R$↘ and $\Delta R$↗, of opposite sign. (b) $2\Delta R_{AHE}$ as a function of the device size. (c) $\Delta R_{AHE}$ as a function of the FM thickness $t_{fm}$, for three GeTe thicknesses. The black line marks $\Delta R_{AHE} = 0$; each curve crosses it at a different t_fm, so the FM thickness can be chosen to cancel the AHE contribution.

**S5 Ferroelectric poling attempt**

The nanofabrication design of the T-shaped devices has the small drawback of having the CoFe shaping with a long wire and step like behavior as explained in the nanofabrication section of the main paper. During the process the exposed surfaces of the stack oxidize on contact with air. The top of the channel is etched immediately before the CoFe deposition, which restores a clean CoFe/Ti/GeTe interface, but the sidewall is not reached by this etch and keeps its native oxide. This oxide should in principle confine the injection to the larger and smoother top interface, but being native and uncontrolled in thickness it is a poor barrier. By having CoFe in direct contact with the entire stack, part of the pulse is shunted, hence reducing its effectiveness and lowering the possibility of switching the ferroelectric state of GeTe. Furthermore, in terms of current density, the smallest section is not at the interface between CoFe and GeTe but the CoFe wire itself, which reduces the maximum applicable voltage before breaking down the nanodevice or observing electromigration phenomena.

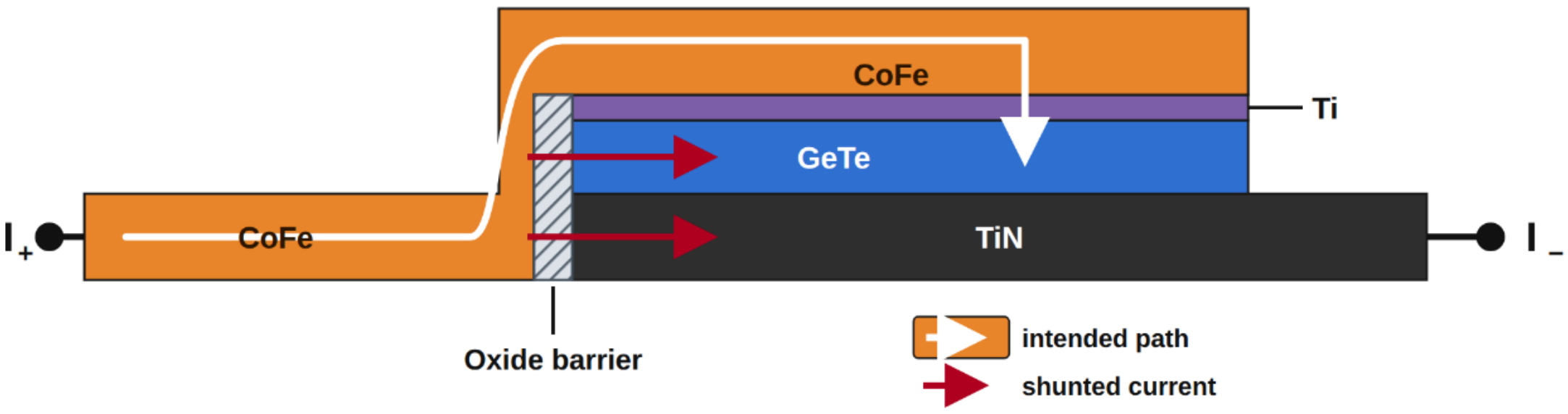

**Figure S8:** Cross-section of the interface between the CoFe electrode and the TiN/GeTe/Ti channel. The CoFe is deposited as a long wire that climbs the step at the edge of the channel. The top of the channel is etched immediately before this deposition, so the current is intended to be injected into GeTe through the clean CoFe/Ti interface (white arrow). The native oxide grown on the sidewall during air exposure is not removed by that etch and remains an uncontrolled, imperfect barrier, so part of the current is shunted directly into GeTe and TiN without crossing the Ti interface (red arrows). Layer colors follow Figure 2.

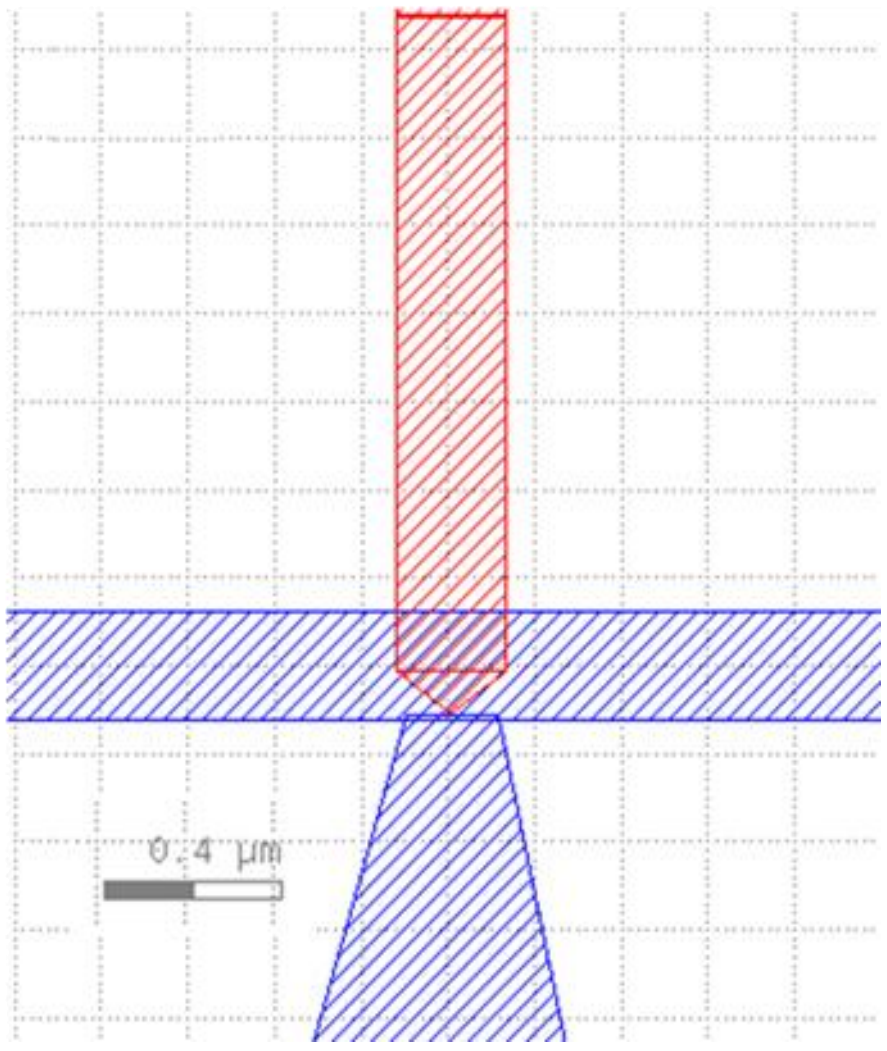

**Figure S9**: KLayout mask design of the spin-orbit read-out devices. The ferromagnetic electrode is drawn in red and the channel in blue; their overlap defines the injection interface.